\documentclass[%
reprint,
superscriptaddress,
amsmath,amssymb,
 aps]{revtex4-2}

\usepackage{graphicx}
\usepackage{bm}
\usepackage{xcolor}
\usepackage{subfigure}

\usepackage{comment}
\newcommand{\bk}{\boldsymbol{k}}
\newcommand{\bq}{\boldsymbol{q}}
\newcommand{\br}{\boldsymbol{r}}

\newcommand{\bK}{\boldsymbol{K}}
\newcommand{\brho}{\boldsymbol{\rho}}
\newcommand{\bQ}{\boldsymbol{Q}}

\begin{document}

\title{Vortical effects in chiral band structures }

\author{Swadeepan Nanda}
\affiliation{University of Houston, Houston, TX}
\author{Pavan Hosur}
\affiliation{University of Houston, Houston, TX}

\date{\today}

\begin{abstract}
The chiral vortical effect is a chiral anomaly-induced transport phenomenon characterized by an axial current in a uniformly rotating chiral fluid. It is well-understood for Weyl fermions in high energy physics, but its realization in condensed matter band structures, including those of Weyl semimetals, has been controversial. In this work, we develop the Kubo response theory for electrons in a general band structure subject to space- and time-dependent rotation or vorticity relative to the background lattice. For continuum Hamiltonians, we recover the chiral vortical effect in the static limit and the transport or uniform limit when the fluid, strictly, is not a Fermi liquid. In the transport limit of a Fermi liquid, we discover a new effect that we dub the gyrotropic vortical effect. The latter is governed by Berry curvature of the occupied bands while the former contains an additional contribution from the magnetic moment of electrons on the Fermi surface. The two vortical effects can be understood as kinematic analogs of the well-known chiral and gyrotropic magnetic effects in chiral band structures. We address recent controversies in the field and conclude by describing device geometries that exploit Ohmic or Seebeck transport to drive the vortical effects.

\end{abstract}

\maketitle

\section{Introduction} Chiral transport phenomena in three dimensions have garnered tremendous interest in condensed matter physics since the discovery of Weyl semimetals (WSMs) -- three dimensional topological materials defined by the presence of accidental intersections between non-degenerate bands \citep{VafekDiracReview,Burkov2018,Burkov:2016aa,YanFelserReview,ArmitageWeylDiracReview,Shen2017,Belopolski:2016wu,Guo2018,Chang2016,Gyenis_2016,Huang:2015vn,Inoue1184,Lv:2015aa,Sun2015a,Xu2015,Xu2016,Xu613,Yang:2015aa,Zheng2016}. Near these intersections or Weyl nodes, the Hamiltonian resembles that of massless, relativistic Weyl fermions. Weyl nodes have a well-defined handedness or chirality, defined as the absence of all improper symmetries such as reflection and inversion ($\mathcal{I}$), and behave like unit magnetic monopoles for Berry flux in momentum space. Chiral transport phenomena in WSMs are characterized by distinct responses of right- and left-handed Weyl fermions to external perturbations such as electromagnetic fields and can invariably be traced to the chiral anomaly, defined as the violation of the classical $U(1)$ chiral gauge symmetry by the quantum path integral \cite{NielsenABJ}. Although the anomaly, first discovered in high-energy physics, is strictly absent in WSMs as they necessarily contain equal numbers of right- and left-handed Weyl nodes \citep{NielsenFermionDoubling1,NielsenFermionDoubling2}, clever ways of resolving the nodes have led to a myriad of anomalous behaviors of WSMs subject to electromagnetic fields  \citep{Hosur2013a,Wang_2018,Hu:2019aa,ZyuninBurkovWeylTheta,ChenAxionResponse,VazifehEMResponse,Burkov_2015,Hosur2012,Juan:2017aa,Wang2017,Nagaosa:2020aa,IsachenkovCME,SadofyevChiralHydroNotes,GoswamiFieldTheory,Wang2013,BasarTriangleAnomaly,LandsteinerAnomaly,Loganayagam2012,PhysRevD.106.045022}.

Fundamentally, the anomaly manifests as a non-conservation of chiral charge in the presence of parallel electric and magnetic fields even though the low energy Weyl Hamiltonian naïvely predicts chiral charge conservation. Alternately, it generates the chiral magnetic effect (CME), defined as an equilibrium, dissipationless current along a constant magnetic field: $\boldsymbol{j}^\text{CME}\propto\boldsymbol{B}$ \citep{Vilenkin1980,KHARZEEV2014133}. Such a current is allowed in the continuum, where purely left-handed (or purely right-handed) Weyl fermions can exist. However, similarly to the anomaly, the CME too is forced to vanish in WSMs due to lattice regularization. The core difference between the continuum and the lattice versions of the CME generated great debate during its adaptation from high-energy to condensed matter physics \cite{Zyunin2012,ChenAxionResponse,goswami2013chiral,VazifehEMResponse,ChangCME2015,ChangCMEb2015,Goswami2015}. The controversies were eventually resolved by extending the response to non-zero frequency ($\omega$) and momentum ($\bq$) and distilling subtleties of the dc limit. The \emph{static} limit ($\omega\to0$ before $q\to0$) leads to a non-zero CME only if the system is held in a non-equilibrium steady state and directly measures the Berry monopole charge enclosed by the occupied states \citep{Zhong2016,Son2012}. In contrast, the \emph{uniform} limit ($q\to0$ before $\omega\to0$) was termed the gyrotropic magnetic effect (GME) \citep{Zhong2016} and was shown to describe the equilibrium response to a time-dependent magnetic field, or equivalently, a circulating electric field: $\boldsymbol{j}^\text{GME}\propto\boldsymbol{\nabla}\times\boldsymbol{E}$. Interestingly, it can exist even in band structures that lack Weyl nodes but break time-reversal ($\mathcal{T}$)
and $\mathcal{I}$ symmetries.

The chirality, however, is an intrinsic property of the Weyl Hamiltonian
that does not rely on coupling to electromagnetic fields. A natural
question is, ``what chiral transport phenomena do neutral Weyl fermions exhibit?'' A striking such response is the chiral vortical effect (CVE), defined as the appearance of a dissipationless axial current in a rotating Weyl fluid, $\boldsymbol{j}^\text{CVE}\propto\boldsymbol{\Omega}$, where $\boldsymbol{\Omega}$ is the angular velocity. It was initially predicted for neutrino fluxes from rotating black holes and other chiral relativistic field theories \citep{Vilenkin1978a, Vilenkin1978b,Vilenkin1979, Vilenkin1980,Vilenkin1980c,Rogachevsky2010,Loganayagam2012,StephanovKineticTheory,Chen2014,Prokhorov2018,Prokhorov2018a,Prokhorov:2019aa,zakharov2018quarkhadron,Stone2018} and has been observed in heavy-ion collisions \citep{Kharzeev:2016vn, Bevan1997}. Historically, it was usually described as a CME where the Coriolis force in a rotating frame simulates $\boldsymbol{B}$. However, it differs from the CME in key ways. Firstly, it is arguably more fundamental as it is a strictly kinematic effect that does not rely on gauge fields. Secondly, it is much less understood in condensed matter; its historical derivations using Boltzmann and hydrodynamic equations have yielded elegant solutions in the static limit, especially for Lorentz invariant Weyl fermions \cite{Rogachevsky2010,Loganayagam2012,StephanovKineticTheory,Chen2014,Prokhorov2018a,Prokhorov2018,zakharov2018quarkhadron,Stone2018,Prokhorov:2019aa,Shitade2020,Toshio2020,KhaidukovCVEFL,KirilinCVESF}, but its generalization to arbitrary band structures at finite $\bq$ and $\omega$ remains an open problem. Thirdly, electrons under a static $\boldsymbol{B}$ are within the purview of Bloch's theorem that forbids an equilibrium current in an infinite system \cite{Bohm1949,Ohashi1996,Yamamoto2015}, but the theorem is inapplicable for a fluid rotating at constant angular velocity. We will return to this point later. Finally, both lattice and continuum realizations of the CME can be triggered in practice by merely applying a suitable $\boldsymbol{B}$. In contrast, a continuous fluid can be physically rotated to trigger the CVE, but crystalline solids do not provide a mechanical handle for rotating the electrons in
their bands. A practical route to rotate the electrons in a solid is essential for
harnessing the phenomenon for device applications.

In this work, we develop the theoretical framework for describing the current response of electrons in a general band structure rotated relative to a stationary lattice at a space- and time-dependent angular velocity $\boldsymbol{\Omega}(\boldsymbol{r},t)$. Alternately, the current can be interpreted as a response to space and time-dependent vorticity $\boldsymbol{\mathcal{V}}(\boldsymbol{r},t) = \boldsymbol{\nabla}\times \boldsymbol{u}(\boldsymbol{r},t)/2$, where $\boldsymbol{u}(\boldsymbol{r},t)$ is the velocity field of the electron fluid. This is because, for smooth and slow dependence on $\boldsymbol{r}$ and $t$, $\boldsymbol{u}(\boldsymbol{r},t) = \boldsymbol{\Omega}(\boldsymbol{r},t)\times\boldsymbol{r}$ immediately yields $\boldsymbol{\mathcal{V}}(\boldsymbol{r},t)=\boldsymbol{\Omega}(\boldsymbol{r},t)$. In other words, the vorticity locally mimics angular velocity about the vortex axis. 

We will focus on the static and uniform limits, which respectively describe the response to time-independent and time-dependent spatially uniform vorticity or angular velocity. The static limit corresponds to the CVE; interestingly, the uniform limit yields the same response function provided the quasiparticle lifetime $\tau$ remains finite as the frequency $\omega\to0$. In contrast, precisely in the uniform or transport limit of a Fermi liquid, defined by $\omega\tau\to\infty$ as $\omega\to0$, we discover a new response that we term the gyrotropic vortical effect (GVE). The GVE can be viewed as an axial current in response to angular acceleration, $\boldsymbol{j}^\text{GVE}\propto\boldsymbol{\alpha}(\br,t)=\partial_t\boldsymbol{\Omega}(\br,t)$, but its gyrotropic nature becomes transparent when viewed as the response to circulating acceleration $\boldsymbol{\nabla}\times\boldsymbol{a}(\br,t)\equiv\boldsymbol{\nabla}\times\partial_t\boldsymbol{u}(\br,t)$. Thus, just as the CVE is usually interpreted as the rotational analog of the CME, the GVE proposed here can be understood as the rotational or vortical counterpart of the GME.

\begin{figure}
\begin{center}
\includegraphics[width=0.4\columnwidth]{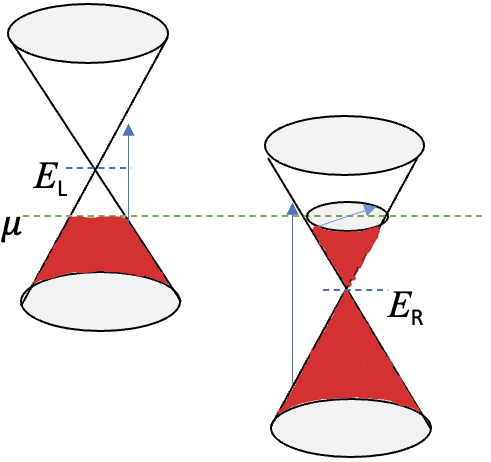}
\caption {Schematic of microscopic processes causing the vortical effects in a WSM with left- and right-handed Weyl nodes at different energies. The GVE arises from non-resonant ($E_{\bk}^m-E_{\bk}^n\neq\omega$) interband ($m\neq n$) virtual processes at $q=0$ (vertical arrows) and survives as $\omega\to0$ provided $E_L\neq E_R$. The CVE appears in the static limit $\omega=0$, $q\to0$, and contains an additional contribution from intraband ($m=n$) resonant ($E_{\bk}^n=E_{\bk+\bq}^n$) processes (horizontal arrow).\label{fig:micro}}
\end{center}
\end{figure}

Microscopically, we show that the GVE stems from purely interband virtual processes in the clean limit whereas the CVE
relies on both interband and intraband processes. In other words, we show that the (isotropic parts of the) relevant linear response functions are
\begin{subequations}
\label{eq:main}
\begin{align}
        \chi^{\text{GVE}}&=\chi^\text{inter}_\text{iso}\label{eq:main2}\\
        \chi^{\text{CVE}}&=\chi^\text{intra}_\text{iso}+\chi^\text{inter}_\text{iso}\label{eq:main1}
\end{align}
\end{subequations}
 Eqs.~(\ref{eq:main}) summarize our main results. Explicit expressions for $\chi^\text{inter}_\text{iso}$ and $\chi^\text{intra}_\text{iso}$ are given later in Eqs.~(\ref{eq:chi-GVE}) and~(\ref{eq:intra}), while Fig.~(\ref{fig:micro}) depicts the corresponding microscopic processes for a pair of Weyl nodes.

\section{Vortical response theory} We now develop the vortical response theory and derive expressions for $\chi^{\text{intra}}$ and $\chi^{\text{inter}}$. We begin by imagining electrons governed by a Bloch Hamiltonian $H_{0}(\bk)$ driven in such a way that they rotate by a space- and time-dependent angular velocity $\boldsymbol{\Omega}(\boldsymbol{r},t)$. This is a non-equilibrium system; to apply the Kubo formalism, we must first recast this problem into one of equilibrium perturbation theory. To that end, we note that rotation induces an additional time dependence into the time evolution of any wavefunction $\psi(\br,t)$:
$\psi(\br,t)=T\left[e^{i\intop_0^t\boldsymbol{L}\cdot\boldsymbol{\Omega}(\br,t')-iH_0(\bk)t}\right]\psi(\br,0)$ in units $\hbar=1$, where $T[\dots]$ denotes time-ordering and $\boldsymbol{L}$ is the spatial angular momentum that generates rotations on length scales large compared to the lattice constant. Unlike usual perturbative calculations where the Hamiltonian is known, and the wavefunction is determined perturbatively, the time dependence of the wavefunction is physically given here, and the perturbation must be inferred. Suggestively defining $\psi_I(\br,t)=e^{iH_0(\bk)t}\psi(\br,t)$, we see that $i\partial_t\psi_I(\br,t)=-\boldsymbol{L}\cdot\boldsymbol{\Omega}(\br,t)\psi_I(\br,t)$. In other words, $\psi_I(\br,t)$ behaves like an interaction picture wavefunction corresponding to an unperturbed Hamiltonian $H_0(\bk)$ and perturbation $-\boldsymbol{L}\cdot\boldsymbol{\Omega}(\br,t)$, implying a total Hamiltonian
\begin{equation}
H(\bk,\boldsymbol{r},t)=H_{0}(\bk)-\boldsymbol{L}\cdot\boldsymbol{\Omega}(\boldsymbol{r},t)\label{eq:H}
\end{equation}
while the appropriate equilibrium many-body state is the equilibrium Fermi sea of $H_0(\bk)$. Now, the full machinery of equilibrium perturbation theory can be deployed. In particular, the Kubo formula can be used to compute the linear current response to the perturbation $-\boldsymbol{L}\cdot\boldsymbol{\Omega}(\boldsymbol{r},t)$.

A few subtleties arise here, however, since $\boldsymbol{L}$ is not
translationally invariant, yet the corresponding vertex conserves
momentum. In the Bloch basis,
\begin{align}
 \langle\psi^{m}_{\bk}|-\boldsymbol{L}\cdot\boldsymbol{\Omega}(\br, t)|\psi^{n}_{\bk+\bq'}\rangle&=(2\pi)^3i\boldsymbol{\nabla}_{\bq'}\delta(\bq'-\bq)\times\nonumber\\
\langle u^{m}_{\bk}|&(\bk-i\boldsymbol{\nabla}_{\br})|u^{n}_{\bk+\bq'}\rangle\cdot\boldsymbol{\Omega}(\bq,t)
\end{align}
where $|\psi_{\bk}^{n}\rangle$ is the Bloch wavefunction in the $n$-th band and $|u_{\bk}^{n}\rangle$ is its periodic part. Thus, an angular velocity field with momentum $\bq$ can produce currents $\boldsymbol{j}(\bq')$ distributed around $\bq$ through the $\delta(\bq'-\bq)$ function. Since measurement probes have a finite resolution, the physical current is given by the integrated weight
\begin{equation}
\left\langle j_{\alpha}(\bq,i\omega_{n})\right\rangle =\intop_{\boldsymbol{q}'}\tilde{\chi}_{\alpha\beta}(\boldsymbol{q}',\boldsymbol{q},i\omega_{n})\Omega_{\beta}(\boldsymbol{q},i\omega_{n})
\end{equation}
as a function of Matsubara frequencies, where
\begin{eqnarray}
\tilde{\chi}_{\alpha\beta}(\boldsymbol{q},\boldsymbol{q}',i\omega_{n})&=&-T\sum_{\bk,i\nu_{n}}\text{tr}\left[j_{\alpha}(\bk+\boldsymbol{q}')G(\bk,i\nu_{n})\times\right.\nonumber\\&&\left. L_{\beta}(\bq';\bq)G(\bk+\boldsymbol{q}',i\nu_{n}+i\omega_{n})\right]    
\end{eqnarray}
and $G(\bk,i\nu_{n})=\left[i\nu_{n}-H_{0}(\bk)+i\text{sgn}(\nu_n)/2\tau\right]^{-1}$
as usual. The physical, retarded response function is given by $\chi_{\alpha\beta}(\boldsymbol{q},\omega)=\intop_{\boldsymbol{q}'}\tilde{\chi}_{\alpha\beta}(\boldsymbol{q},\boldsymbol{q}',i\omega_{n}\to\omega+i0^{+})$ for which we provide an explicit expression in App. \ref{App:general results} that is valid at arbitrary $T$, $\omega$ and $\bq$ and for general lattice models.

\begin{figure}
    
\subfigure[]{\includegraphics[width=0.3\columnwidth]{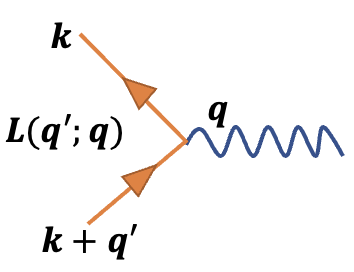}} \subfigure[]{\includegraphics[width=0.5\columnwidth]{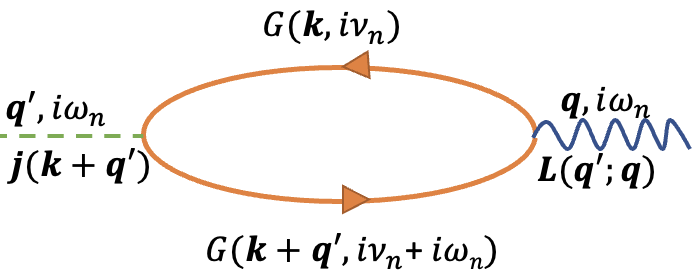}}\caption{(a) The angular momentum vertex. (b) The first-order Feynman diagram for the response.}
\end{figure}

\section{Results}

 Henceforth, we focus on the isotropic part of the response, $\chi_\text{iso}(\bq,\omega)=\frac{1}{3}\chi_{\alpha\alpha}(\bq,\omega)$. It is useful to separate it into interband and intraband terms, 
 $\chi_\text{iso}$:
\begin{equation}
    \chi_\text{iso} = \chi_\text{iso}^\text{inter} + \chi_\text{iso}^\text{intra}
\end{equation}
Some algebra, described in App. \ref{App:angular momentum}, leads to
 \begin{align}
     \chi_\text{iso}^\text{inter}=&\frac{i}{3}\intop_{\bk}\sum_{n\neq m}\langle u_{\bk}^{m}|\bQ_{\bk}| u_{\bk+\bq}^{n}\rangle\cdot\nonumber\\&\langle u^{m}_{\bk}|\boldsymbol{\nabla}_{\bk}u^{n}_{\bk}\rangle \times\langle u^{n}_{\bk}|\boldsymbol{j}(\bk)|u^{m}_{\bk}\rangle S_{nm}(\bk)\nonumber\\
    \chi_\text{iso}^\text{intra}=&\frac{i}{3}\intop_{\bk}\sum_{n}\langle u_{\bk}^{m}|\bQ_{\bk}| u_{\bk+\bq}^{n}\rangle\cdot\nonumber\\&\langle u^{n}_{\bk}|\left(\boldsymbol{j}_{\bk}^n-\boldsymbol{v}_{\bk}\right)\times|\boldsymbol{\nabla}_{\bk}u^{n}_{\bk}\rangle S_{nn}(\bk)\label{eq:main}
 \end{align}
where
\begin{equation}
    S_{nm}(\bk)=\intop_{\nu}\text{Im}\left[\frac{2}{\nu+\frac{i}{2\tau}}\right]\frac{f(E^m_{\bk}+\nu)-f(E_{\bk+\bq}^n-\nu)}{\nu+E^m_{\bk}-E_{\bk+\bq}^n+\omega+\frac{i}{2\tau}}\label{eq:Snm}
\end{equation}
evaluated in the desired limit of $q$, $\omega$, $\tau$. Here, $f(\nu)$ is the usual Fermi function that reduces to a step function at $T=0$. 

A peculiar quantity in the above expressions is $\bQ_{\bk} = \bk-i\boldsymbol{\nabla_{\brho}}$, a momentum-like object containing two parts: the crystal momentum $\bk$ and spatial derivative within the unit cell $i\boldsymbol{\nabla}_{\brho}$. Together, they respect the periodicity of the Brillouin zone. Alternately, $\bQ_{\bk}$ can be viewed as a gauge invariant momentum corresponding to the ``gauge symmetry" $\bk\to\bk+\bK$, where $\bK$ is a reciprocal lattice vector, and ``gauge field" $\boldsymbol{\mathcal{A}}_{\bk}^{mn}(\bq')=i\langle u^{n}_{\bk}|\boldsymbol{\nabla}_{\brho}|u^{m}_{\bk+\bq'}\rangle$ that captures the freedom in choosing the Brillouin zone. In certain physical regimes described in App. \ref{App:continuum}, such as the nearly free electron limit with $\bk$ far from the Brillouin zone edges, $\bQ_{\bk}$ reduces to the usual continuum momentum $\bk$. 

$\bQ_{\bk}$ appears in our expressions because we are calculating the response to angular velocity, which couples to spatial angular momentum. $\bQ$ is analogous to the lattice charge current that couples and captures the response to Peierls electromagnetic fields. In other words, just as continuum current is well-defined independent of any lattice but acquires interesting structure and inherits the lattice periodicity when projected onto Bloch states, the continuum momentum should be replaced by the $\bk$-periodic quantity $\bQ_{\bk}$ that contains contributions from Bloch functions.




Eqs. (\ref{eq:main}) and (\ref{eq:Snm}) describe the response of Bloch electrons in a general lattice to rotation at arbitrary $\bq$ and $\omega$. To proceed further analytically, we (i) set $T=0$, (ii) take the continuum limit and replace $\bQ_{\bk}\to\bk$ (iii) assume no band intersections near the Fermi level, and (iv) focus on the regime of small of $q$, $\omega$ and $1/\tau$. Although $\tau$ was introduced phenomenologically, we allow it to have an implicit dependence on $\bq$ and $\omega$. Thus, we study the uniform limit, $q=0$ followed by $\omega\to0$, in two regimes: $|\omega\tau|\to0$ and $|\omega\tau|\to\infty$, the latter defining a Fermi liquid due to a diverging quasiparticle lifetime as $\omega\to0$. Similarly, we investigate the static limit, $\omega=0$ followed by $q\to0$, in the two regimes $q\tau\to0$ and $q\tau\to\infty$. We emphasize that the large $\tau$ regimes of the uniform and static limits include the cases where $\tau\to+\infty$ from the outset. Thus, the six permutations for the orders of $q$, $\omega$ and $1/\tau$ define four distinct physical regimes. We now provide results for the interband and intraband contributions to $\chi_\text{iso}$ in the limits $q\to0$, $\omega\to0$ and $1/\tau\to0$ in each of the four regimes.

\subsection{GVE in the uniform, Fermi liquid limit}

   \begin{table*}
   \centering
   \begin{tabular}{c|c|c|c}
   \hline
    Limit & Definition & $\chi_\text{iso}^\text{inter}$ & $\chi_\text{iso}^\text{intra}$\\
    \hline
    \hline
    Uniform, clean & $vq\ll|\omega|$, $1/\tau\ll|\omega|$, arbitrary $vq\tau$ & $-\frac{2}{3}\intop_{\bk}\sum_n \Theta(-E_{\bk}^n)\bk\cdot\boldsymbol{F}_{\bk}^n$ & 0\\[3pt]
    \hline
    Uniform, dirty & $vq\ll|\omega|\ll1/\tau$ & $-\frac{2}{3}\intop_{\bk}\sum_n \Theta(-E_{\bk}^n)\bk\cdot\boldsymbol{F}_{\bk}^n$ & $-\frac{2}{3}\intop_{\bk}\sum_{n} \frac{\boldsymbol{m}_{\bk}^n\cdot\bk}{e}\delta(E_{\bk}^n)$\\[3pt]
    \hline
    Static, clean & $|\omega|\ll vq$, $1/\tau\ll vq$, arbitrary $\omega\tau$ & $-\frac{2}{3}\intop_{\bk}\sum_n \Theta(-E_{\bk}^n)\bk\cdot\boldsymbol{F}_{\bk}^n$ & $-\frac{2}{3}\intop_{\bk}\sum_{n} \frac{\boldsymbol{m}_{\bk}^n\cdot\bk}{e}\delta(E_{\bk}^n)$\\[3pt]
    \hline
    Static, dirty & $|\omega|\ll vq \ll 1/\tau$ & $-\frac{2}{3}\intop_{\bk}\sum_n \Theta(-E_{\bk}^n)\bk\cdot\boldsymbol{F}_{\bk}^n$ & $-\frac{2}{3}\intop_{\bk}\sum_{n} \frac{\boldsymbol{m}_{\bk}^n\cdot\bk}{e}\delta(E_{\bk}^n)$\\[3pt]
    \hline

\end{tabular}
\caption{Summary of results at $q\to0$, $\omega\to0$ and $1/\tau\to0$ in various orders. $v$ is a typical band velocity. The interband susceptibility is the same in all cases. In contrast, the intraband term vanishes in the uniform, clean limit but is the same in all other limits.}
\label{tab:summary-of-results}
\end{table*}

  At $q=0$ and in the Fermi liquid limit, defined by long-lived quasiparticles with lifetime $\tau\gg1/|\omega|$ as $\omega\to0$, the intraband term $S_{nn}(\bk)\to0$ while the interband terms are non-zero. Some algebraic manipulations yield an expression
\begin{equation}
\chi^\text{inter}_\text{iso}=-\frac{2}{3}\sum_{n}\intop_{\bk}\Theta(-E_{\bk}^{n})\boldsymbol{F}_{\bk}^{n}\cdot{\bk}
\label{eq:chi-GVE}
\end{equation}
where $\boldsymbol{F}_{\bk}^{n}=i\left\langle \boldsymbol{\nabla}_{\bk}u_{\bk}^{n}\left|\times\right|\boldsymbol{\nabla}_{\bk}u_{\bk}^{n}\right\rangle $
is the Berry curvature of the $n^{th}$ band. The term itself is insensitive to the order in which $q$, $\omega$ and $1/\tau$ were taken to zero, and is the only one that survives in the Fermi liquid limit defined above. In analogy with the GME, we term this effect the GVE.

The GVE, defined completely by $\chi_{\text{iso}}^{\text{inter}}$ according to Eq. (\ref{eq:main2}), is purely real, which implies a dissipationless current that arises purely from interband processes between non-degenerate Bloch states. It vanishes in systems with an improper symmetry. For instance, a mirror plane normal to $z$ transforms $k_z\to-k_z$, $k_{(x,y)}\to k_{(x,y)}$ and $F^n_{{\bk}z}\to F^n_{{\bk}z}$, $F^n_{{\bk}(x,y)}\to -F^n_{{\bk}(x,y)}$, while $\mathcal{I}$ transforms $\bk\to-\bk$ and $F^n_{{\bk}}\to F^n_{{\bk}}$. However, it is generically non-zero in chiral, $\mathcal{T}$-symmetric
systems, which obey $\boldsymbol{F}_{\bk}^{n}=-\boldsymbol{F}_{-\bk}^{n}$. 
 Naïvely, $\chi^\text{inter}_\text{iso}$ seems poorly regularized
as it receives contributions from all occupied states. However, the
total contribution from a filled band vanishes at $T=0$ in the continuum
limit, when the Brillouin zone can be compactified to a sphere
and all points at $k\to\infty$ are identified. Specifically, $\intop_{BZ}\boldsymbol{F}_{\bk}^{n}\cdot\bk=\intop_{\partial BZ}(\boldsymbol{A}_{\bk}^{nn}\times\bk)\cdot d\boldsymbol{s}$
using Gauss's divergence theorem, but the last expression vanishes since $|u_{\bk}^{n}\rangle$
must be constant on $\partial BZ$ due to the above boundary condition.
Thus, only partially occupied bands contribute to $\chi^\text{inter}_\text{iso}$.


\subsection{CVE in all other limits}

In contrast to $\chi_\text{iso}^\text{inter}$, $\chi_\text{iso}^\text{intra}$ depends strongly on the order of limits. It vanishes in the Fermi liquid limit, while the other three limits lead to
\begin{equation}
\chi^\text{intra}_\text{iso} =-\frac{2}{3}\sum_{n}\intop_{\bk}\delta(E_{\bk}^{n})\frac{\bk\cdot \boldsymbol{m}_{\bk}^n}{e} \label{eq:intra}
\end{equation}
where $\boldsymbol{m}^n_{\bk} = \frac{ie}{2}\langle\boldsymbol{\nabla}u^n_{\bk}|\times(H_{\bk} - E^n_{\bk})|\boldsymbol{\nabla}u^n_{\bk}\rangle$ denotes the orbital moment of the Bloch state $|u_{\bk}^n\rangle$. Similar to $\chi^\text{inter}_\text{iso}$, $\chi^\text{intra}_\text{iso}$ is also purely real and hence generates a dissipationless current. Moreover, $\boldsymbol{m}^n_{\bk}$ transforms the same way as $\boldsymbol{F}^n_{\bk}$ under symmetry operations, so $\chi^\text{intra}_\text{iso}$, similarly to $\chi^\text{inter}_\text{iso}$, is generically non-zero for chiral band structures. The results for $\chi_\text{iso}^\text{inter}$, $\chi_\text{iso}^\text{intra}$ and $\chi_\text{iso}$ are summarized in Table. \ref{tab:summary-of-results}.

The static, clean limit, defined as $\omega\to0$ followed by $q\to0$ with $vq\tau\to\infty$ where $v$ is a typical band velocity, has been well-studied and referred to as the CVE. Interestingly, we find that various other limits also yield susceptibilities that match the CVE susceptibility since the latter is given by the sum of $\chi^\text{intra}_\text{iso}$ and $\chi^\text{inter}_\text{iso}$ according to Eq. (\ref{eq:main1}). Thus, the CVE susceptibility should be easier to observe in experiments as it is robust to the order of limits as long as we are not in the Fermi liquid transport regime.

\subsection{Application to Weyl fermions}

\begin{table*}[]
    \centering
    \begin{tabular}{c|c|c|c|c}
    \hline
    Limit & Definition & $\chi_\text{Weyl}^\text{inter}$ & $\chi_\text{Weyl}^\text{intra}$ & $\chi_\text{Weyl}$\\
    \hline
    \hline
    Uniform, clean & $vq\ll|\omega|$, $1/\tau\ll|\omega|$, arbitrary $vq\tau$ & $\frac{C}{3}\chi_0$ & 0 & $\frac{C}{3}\chi_0$\\[3pt]
    \hline
    Uniform, dirty & $vq\ll|\omega|\ll1/\tau$ & $\frac{C}{3}\chi_0$ & $\frac{2C}{3}\chi_0$ & $C\chi_0$\\[3pt]
    \hline
    Static, clean & $|\omega|\ll vq$, $1/\tau\ll vq$, arbitrary $\omega\tau$ & $\frac{C}{3}\chi_0$ & $\frac{2C}{3}\chi_0$ & $C\chi_0$\\[3pt]
    \hline
    Static, dirty & $|\omega|\ll vq \ll 1/\tau$ & $\frac{C}{3}\chi_0$ & $\frac{2C}{3}\chi_0$ & $C\chi_0$\\[3pt]
    \hline

    \end{tabular}
    \caption{Summary of results at $q\to0$, $\omega\to0$ and $1/\tau\to0$ in various orders for an isotropic Weyl fermion with velocity $v$, chiral charge $C$ and chemical potential $\mu$ relative to the Weyl node. Here, $\chi_0=\left(\frac{\mu}{2\pi v}\right)^2$.}
    \label{tab:Weyl-results}
\end{table*}

We now evaluate the susceptibilities for an isotropic Weyl fermion of chirality $C=\pm1$ described by $H_{0}(\bk)=Cv\bk\cdot\boldsymbol{\tau}-\mu$,
where $\boldsymbol{\tau}$ are Pauli matrices. Its energies, Berry
curvatures and magnetic moment for the $n=\pm1$ bands are $E_{\bk}^{n}=nvk-\mu$, $\boldsymbol{F}_{\bk}^{n}=-Cn\frac{\hat{\bk}}{2k^{2}}$ and $\boldsymbol{m}_{\bk}^{n} = -eCn\frac{\hat{\bk}}{2k}$. Thus, for a WSM with Weyl nodes of chirality $C_i$ and velocity $v_i$ at energy $E_i$, $i=1\dots 2N$, and chemical potential $\mu$, Eq. (\ref{eq:chi-GVE}) reduces at $T=0$ to

\begin{equation}
\chi_{\text{WSM}}^{\text{GVE}}=\frac{1}{3}\sum_{i=1}^{2N}C_i\left(\frac{\mu-E_i}{2\pi v_{i}}\right)^2
\end{equation}
In deriving the above, we have subtracted the contribution of undoped Weyl nodes since filled bands do not contribute, as argued earlier. Similarly, 
\begin{align}
\chi^\mathrm{intra}_\text{WSM}&=\frac{2}{3}\sum_{i=1}^{2N}C_i\left(\frac{\mu-E_i}{2\pi v_i}\right)^2
\label{eq:Weyl-intra}
\end{align}
Including $\chi^\text{inter}_\text{iso}\equiv\chi^\text{GVE}_\text{WSM}$, we obtain the CVE for WSMs,
\begin{align}
\chi^\mathrm{CVE}_\text{WSM}&=\sum_{i=1}^{2N}C_i\left(\frac{\mu-E_i}{2\pi v_i}\right)^2
\label{eq:Weyl-CVE}
\end{align}
Eq.~(\ref{eq:Weyl-CVE}) is the well-known expression for the CVE in relativistic Weyl fermions \cite{Rogachevsky2010,Loganayagam2012,StephanovKineticTheory,Chen2014,Prokhorov2018a,Prokhorov2018,zakharov2018quarkhadron,Stone2018,Prokhorov:2019aa}. 

The results for a single Weyl node are summarized in Table \ref{tab:Weyl-results}. While the effects are non-zero for individual Weyl nodes, they vanish in a WSM unless all improper symmetries are broken. If $\mathcal{I}$ or any improper symmetry is present, the chiralities $C_i$ would appear in equal and opposite pairs while the energies $E_i$ and speeds $v_i$ would be equal for nodes within each such pair.

\section{Comparisons} The past decade has seen growing interest in vortical effects in both relativistic and non-relativistic chiral fermions as well as in their magnetic counterparts. We now contrast our approach and results with the existing ones in each context.

Ref. \cite{Chen2014} calculated the CVE in Lorentz invariant kinetic theories and showed that 
\begin{equation}
    \boldsymbol{j}^{\mathrm{CVE}}=\left(\frac{\mu}{2\pi v}\right)^2\boldsymbol{\Omega}
\end{equation}
at $T=0$ for a single right-handed isotropic Weyl fermion, where $v$ is the Weyl velocity and $\mu$ is measured relative to the Weyl node energy. While this result is well-known, Ref. \cite{Chen2014} showed that $\boldsymbol{j}^{\mathrm{CVE}}$ consists of two parts: a Liouville current $\boldsymbol{j}^{\mathrm{CVE}}_I=\frac{1}{3}\boldsymbol{j}^{\mathrm{CVE}}$ from the group velocity of all occupied (unoccupied) states above (below) the Weyl node and a magnetization current $\boldsymbol{j}^{\mathrm{CVE}}_{II}=\frac{2}{3}\boldsymbol{j}^{\mathrm{CVE}}$ from the magnetization of states near the Fermi surface. Moreover, the above separation of terms was shown to arise purely from Lorentz invariance. Interestingly, our results for a single right-handed Weyl fermion show precisely the same separation between interband and intraband terms. In particular, interband processes contribute to the Liouville current while intraband processes determine the magnetization current. Both terms exist in the static and dirty, uniform limits and yield the CVE. In contrast, the clean, uniform limit gives rise to the GVE and receives contributions from interband processes only, thus consisting purely of a Liouville current. Our Kubo formula approach provides microscopic insight into the CVE that complements the arguments in Ref. \cite{Chen2014} based on Lorentz invariance. 

For non-relativistic systems, recent works have triggered a debate regarding the correct description of vortical effects. Ref. \cite{Toshio2020} described hydrodynamic transport in non-centrosymmetric materials and included the effect of vorticity through a coupling $-\boldsymbol{L}\cdot\boldsymbol{\mathcal{V}}$ to the spatial angular momentum. In contrast, Ref. \cite{Shitade2020} also included a spin-vorticity coupling, $-\boldsymbol{S}\cdot\boldsymbol{\mathcal{V}}$. In writing Eq.~(\ref{eq:H}), we have effectively adopted the former approach. 
Physically, this choice describes a fluid rotating relative to a stationary lattice in such a way that its internal degrees of freedom such as spin and orbital do not directly couple to the external force driving the rotation. For instance, for flow driven by an electric field, spin-vorticity coupling is negligible because it is suppressed by the ratio of the fluid speed to the speed of light. On the other hand, if the crystal is mechanically rotated, the approach of Ref. \cite{Shitade2020} is appropriate as rotation of the underlying lattice couples to all degrees of freedom of the electron gas that lives in it. If spin-vorticity coupling is neglected, our results match with those of Ref. \cite{Shitade2020} provided our uniform and static limits are identified with their transport and magnetization currents, respectively, with an extra minus sign for the latter because we calculate the paramagnetic current whereas orbital magnetization current is diamagnetic.

Finally, we compare and contrast the results for the vortical effects with their magnetic counterparts. The magnetic and vortical effects all require broken $\mathcal{I}$ as well as broken $\mathcal{TI}$ symmetries. However, the CME explicitly requires the presence of Weyl nodes while the GME as well as both the vortical effects can exist for general chiral band structures devoid of Weyl nodes. The microscopic contributions to the various effects are also different: the CME and the GME are governed by the net chirality of the occupied bands and the Fermi surface magnetization, respectively. In contrast, the GVE depends on the Berry curvature of the occupied bands whereas the CVE contains additional contribution from the orbital magnetization of states on the Fermi surface.

\section{Implications for Bloch's theorem} Bloch's theorem states that the current density vanishes in the thermodynamic limit in an arbitrary system at equilibrium. In WSMs, it manifests as a vanishing equilibrium CME. To activate a CME, chiral charge first needs to be pumped across Weyl nodes to create a non-equilibrium steady-state. It is often stated that the CVE evades Bloch's theorem because the thermodynamic limit here violates causality. Specifically, if a fluid rotates at constant $\boldsymbol{\Omega}$, particles far enough from the rotation axis will be forced to travel faster than light, so the system must necessarily have a finite size. However, a recent refinement of Bloch's theorem for finite systems shows that the current density is bounded only by the inverse system size in the current direction, i.e., $|\langle j_z\rangle|<O(1/l_z)$ \cite{Watanabe:2019aa}. Therefore, a generic equilibrium system must have vanishing $|\langle j_z\rangle|$ as $l_z\to\infty$ regardless of its transverse dimensions. This contradicts Eq.~(\ref{eq:Weyl-CVE}), which clearly predicts a non-zero CVE.

The resolution to the paradox can be understood in two equivalent ways. From the lab perspective, the energy of a fluid rotating at constant $\boldsymbol{\Omega}$ can always be lowered by slowing down its rotation, whereas Bloch's theorem assumes that the fluid is already in its lowest energy state. Alternately, the rotating frame Hamiltonian in Eq.~(\ref{eq:H}) violates a key assumption of Bloch's theorem, namely, an energy spectrum that is bounded below, since $\boldsymbol{L}$ is unbounded. Specifically, the upper bound on $|\boldsymbol{L}|=|\br\times\bk|$ is of order $l_\perp/a_\perp$, where $a_\perp$ is the lattice constant in the $i$-direction, which diverges in both continuum ($a_\perp\to0$) and thermodynamic ($l_\perp\to\infty$) limits in the transverse directions.

\begin{figure}
\subfigure[]{\includegraphics[width=0.23\textwidth]{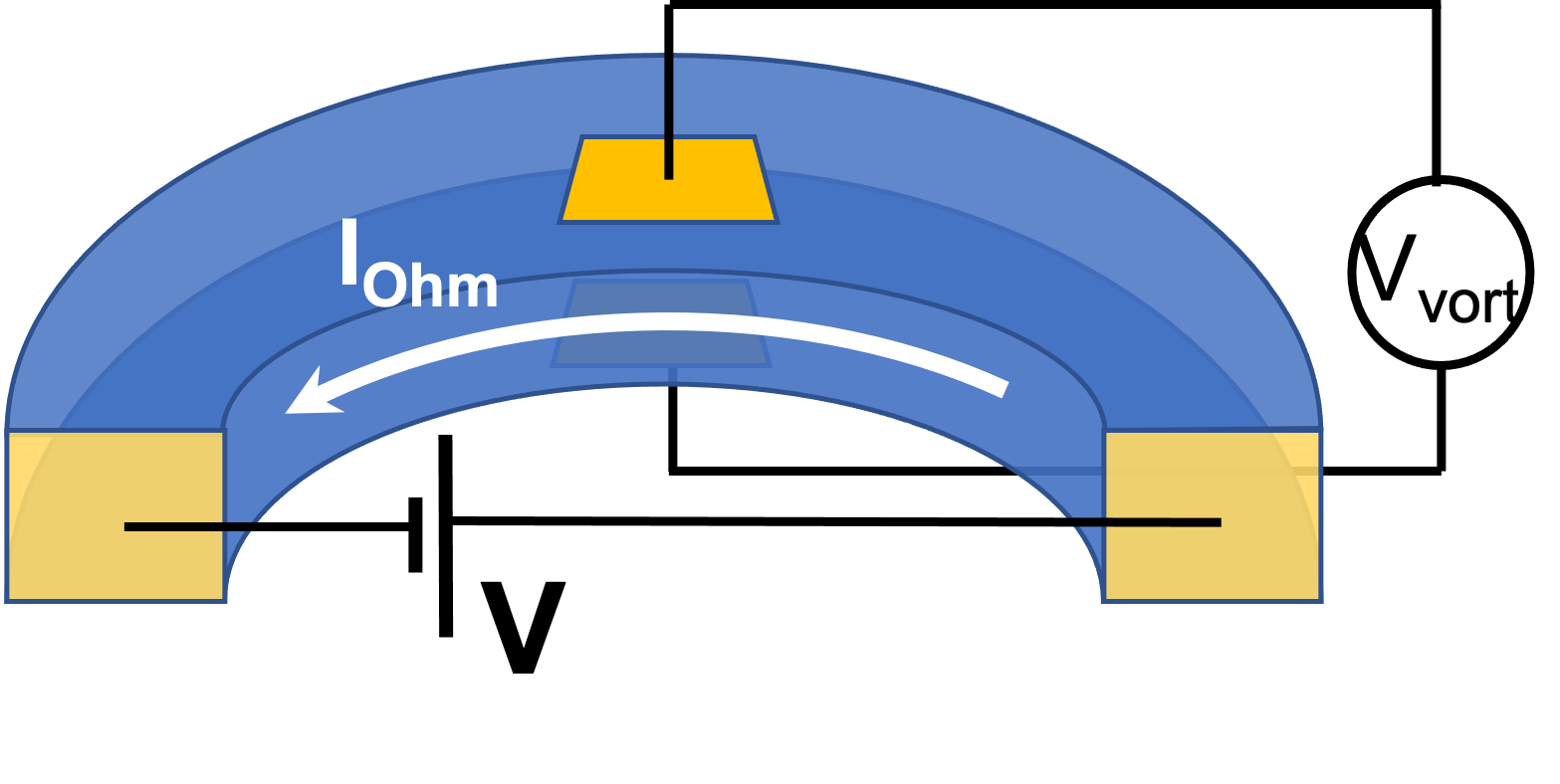}}\subfigure[]{\includegraphics[width=0.23\textwidth]{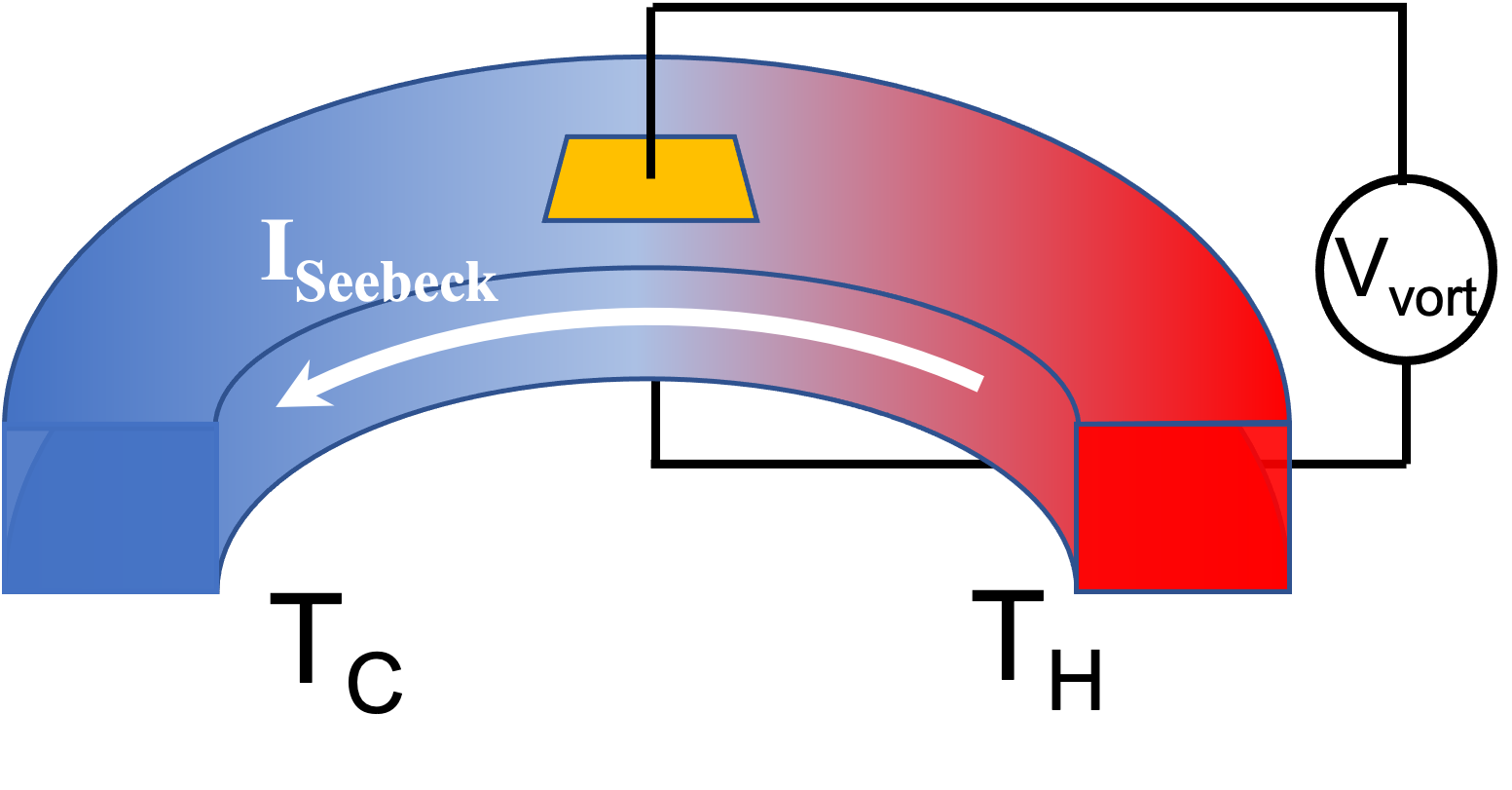}}

\subfigure[]{\includegraphics[width=0.23\textwidth]{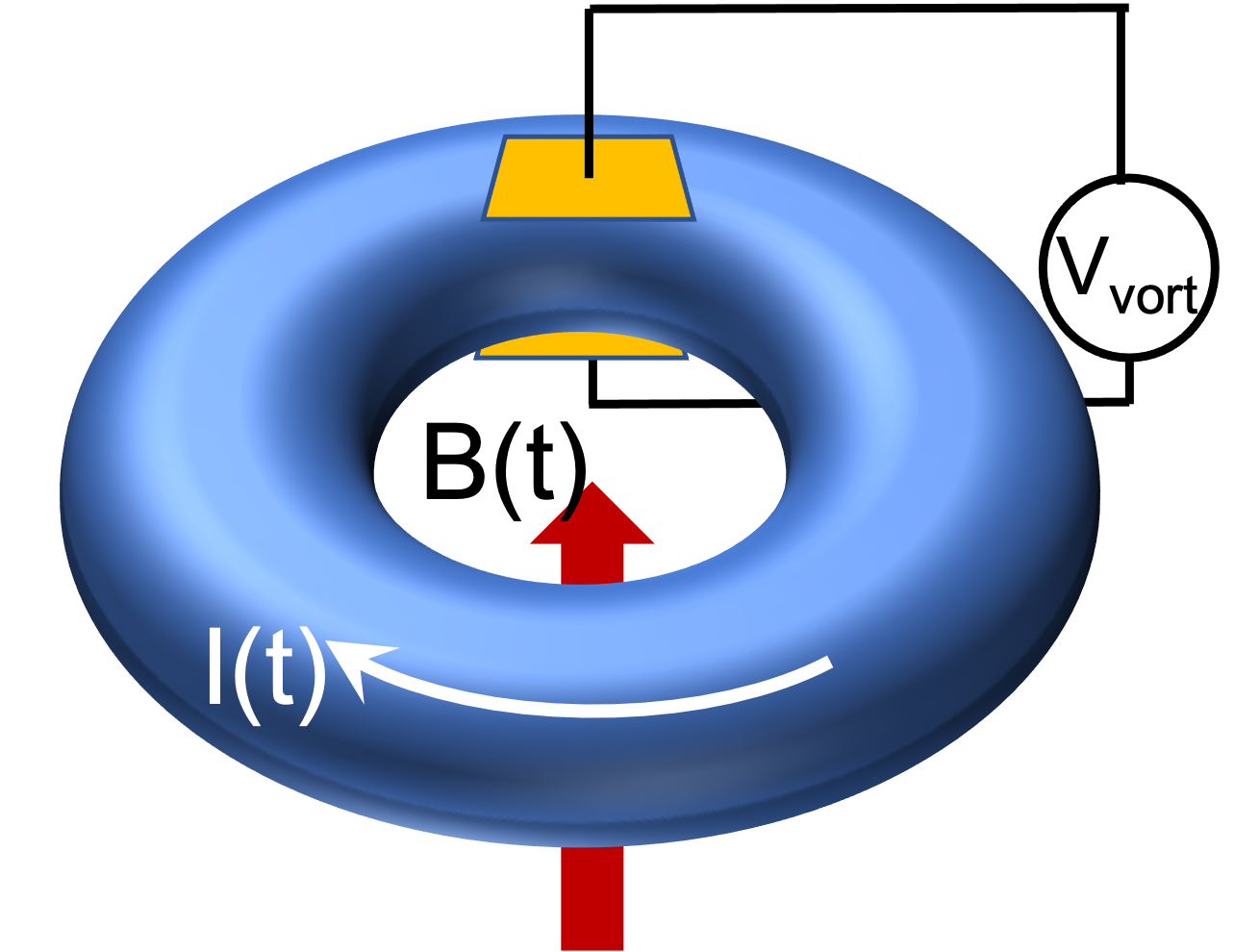}}\subfigure[]{\includegraphics[width=0.24\textwidth]{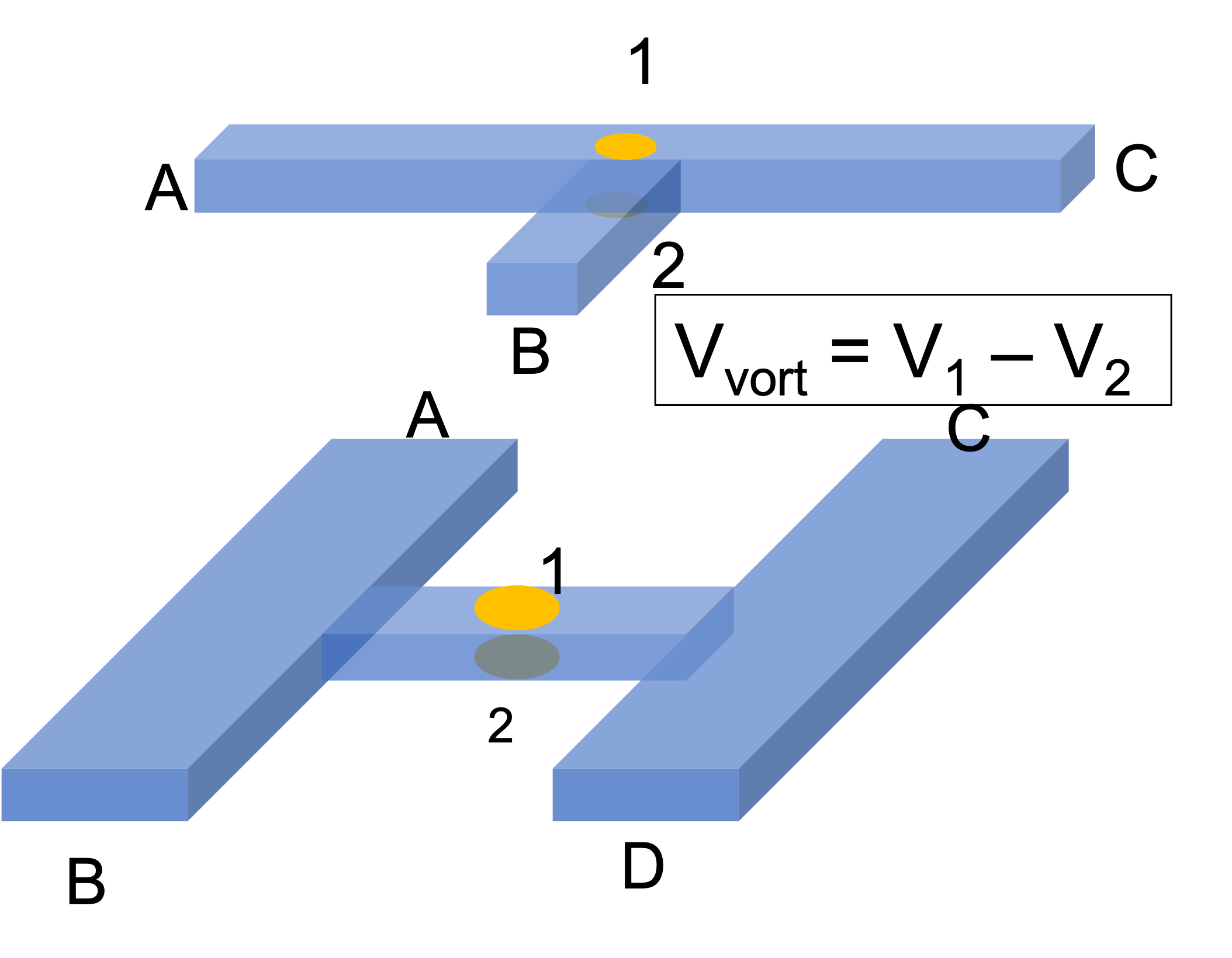}}

\caption{\label{fig:CVE/GVE geometries}Device geometries for utilizing the vortical effects. In (a), (b) and (c), respectively, a current is driven along a curved path by a voltage, a temperature gradient and a time-dependent magnetic field, which produces a vertical potential difference due to the vortical effects. (c) is equivalent to the GME, but here we adopt the perspective that $B(t)$ triggers fluid rotation. In (d, top), $V_{\mathrm{vort}}\protect\neq0$ if a current is driven between $AB$ or $AC$ because it is forced to go around a corner, but $V_{\mathrm{vort}}=0$ for a current between $AC$. (d, bottom) shows a more complex geometry with more terminals and corners that offers richer manipulation. Geometries in (d) can serve as building blocks for scalable circuits.}
\end{figure}

\section{Device geometries} We close by sketching various device geometries that rely on the vortical effects. In these devices, carriers are forced to traverse curved paths by a combination of device geometry and electromagnetic or thermal fields
and hence, endowed with a non-zero $\boldsymbol{\Omega}$
relative to the background lattice. As a result, they develop a voltage
perpendicular to the plane of motion. Figs.~\ref{fig:CVE/GVE geometries}(a)-(c) illustrate the basic ideas for vortical effects driven by an electric
field, thermal gradient and time-dependent magnetic flux, respectively.
If the mobility is $\mu_\text{mob}$ and the
radius of the curved path is $R$, an electric field $E$ induces
a drift velocity $v_{d}=\mu_\text{mob}E$ and hence, $\Omega=\mu_\text{mob}E/R$. Assuming typical values for a WSM, $E=0.1V/m$, $R=1\mu m$, $\mu_\text{mob}=10^{5}cm^{2}V/s$,
$v_{F}=10^{5}m/s$, $(\mu_{+}+\mu_{-})/2=0.5eV$, $\mu_{+}-\mu_{-}=50meV$,
where $\mu_{\pm}$ is the Fermi energy relative to the right/left-handed
Weyl node, we get a large vortical current density of $j=100mA/mm^{2}$. For $q\sim 1/R$, the above numbers result in the CVE (GVE) for $\omega\ll10^{11}$ Hz ($\omega\gg10^{11}$ Hz) 
Fig.~\ref{fig:CVE/GVE geometries}(b) depicts a chiral Nernst effect,
defined as $\boldsymbol{j}(\boldsymbol{r},t)\propto\boldsymbol{\nabla}\times\boldsymbol{\nabla}T(\boldsymbol{r})$.
Note, $\boldsymbol{\nabla}\times\boldsymbol{\nabla}T$ is guaranteed to vanish only if $T(\boldsymbol{r})$ is twice continuously differentiable on all space -- a condition violated by the geometry of Fig. \ref{fig:CVE/GVE geometries}(b). Once again, we expect a large effect in WSMs owing to their large mobility. For $\nabla T\sim 1K/\mu m$ and Seebeck coefficient $S=100\mu V/K$, the resultant electric field is $E=0.1V/m$, which leads to $j=100mA/mm^{2}$. Fig.~\ref{fig:CVE/GVE geometries}(d) shows examples of geometries that can form building blocks of larger circuits. Here, the vortical effects are invoked when the carriers have to turn a corner on their way from a source to a drain.

\acknowledgments
We thank the Department of Energy for funding this work via grant number DE-SC0022264. We are grateful to Shun-Chiao Chang, Zhao Huang, Daniel Bulmash, Karl Landsteiner and Liangzi Deng for insightful discussions.

\onecolumngrid
\appendix

\section{Angular momentum on a lattice}\label{App:angular momentum}
In the continuum, basic quantum mechanics dictates $\hat{\boldsymbol{L}}=\hat{\br}\times\hat{\boldsymbol{p}}$, where $\left<\br|\hat{\boldsymbol{p}}|\br'\right>=-i\boldsymbol{\nabla}_{\br} \delta(\br-\br')$ and we are using hats to distinguish operators from states. In lattice systems such as a rotating optical lattice, one typically recasts $\hat{\boldsymbol{L}}$ into a tight-binding hopping operator by evaluating its matrix elements in the basis of Wannier orbitals. Since we need to evaluate a Kubo formula, which is easier in Fourier space, we project the perturbation $-\hat{\boldsymbol{L}}\cdot\boldsymbol{\Omega}(\hat{\br},t)$ onto Bloch states instead.
\subsection{Projection onto Bloch states}
The Bloch wavefunction for the $n$-th band is generically of the form $\psi^{n}_{\bk}(\br)\equiv\psi^{n}_{\bk}(\boldsymbol{R}+\brho)=N^{-1/2}e^{i\bk\cdot(\boldsymbol{R}+\brho)}u^{n}_{\bk}(\brho)$, where $N$ is the number of unit cells, $\boldsymbol{R}$ is a discrete index that labels them, $\brho$ denotes position within a unit cell, and $u^{n}_{\bk}(\brho)$ is periodic in $\br$ with the same periodicity as the underlying Hamiltonian.
In this basis,
\begin{eqnarray}
   \langle\psi^{m}_{\bk}|-\hat{\boldsymbol{L}}\cdot\boldsymbol{\Omega}(\hat{\br}, t)|\psi^{n}_{\bk+\bq'}\rangle =\intop_{\br,\br'} \psi^{m*}_{\bk}(\br)\br \times i\boldsymbol{\nabla}_{\br}\delta(\br-\br') \psi^{n}_{\bk+\bq'}(\br')\cdot\boldsymbol{\Omega}({\br'},t)
\end{eqnarray}
Suppose $\boldsymbol{\Omega}(\br',t)=e^{-i\bq\cdot\br'}\boldsymbol{\Omega}(\bq)$ is monotonic in space. Approximating $\br\sim\boldsymbol{R}$ and  $\boldsymbol{\nabla}_{\br}=\boldsymbol{\nabla}_{\brho}$, the matrix element becomes

\begin{eqnarray}
   &&\langle\psi^{m}_{\bk}|-\hat{\boldsymbol{L}}\cdot\boldsymbol{\Omega}(\hat{\br}, t)|\psi^{n}_{\bk+\bq'}\rangle = \frac{1}{N}\sum_{\boldsymbol{R}}e^{i(\bq'-\bq)\cdot\boldsymbol{R}} \intop_{\brho,\brho'}e^{-i\bk\cdot\brho+i(\bk+\bq'-\bq)\cdot\brho'}\left[u^{m*}_{\bk}(\brho)\boldsymbol{R}\times i\boldsymbol{\nabla}_{\brho}\delta(\brho-\brho')u^{n}_{\bk+\bq'}(\brho')\right]\cdot\boldsymbol{\Omega}(\bq,t) \nonumber\\
   &=& -\frac{(2\pi)^3}{N}\sum_{\bK}i\boldsymbol{\nabla}_{\bq'}\delta(\bq'-\bq+\bK)\times\left[ \intop_{\brho,\brho'}e^{-i\bk\cdot\brho+i(\bk+\bq'-\bq)\cdot\brho'}u^{m*}_{\bk}(\brho)i\boldsymbol{\nabla}_{\brho}\delta(\brho-\brho')u^{n}_{\bk+\bq'}(\brho')\right]\cdot\boldsymbol{\Omega}(\bq,t)
  \end{eqnarray}
 
where $\bK$ are reciprocal lattice vectors. Integrating by parts over $\brho$ and integrating over $\brho'$,

  \begin{equation}
\langle\psi^{m}_{\bk}|-\hat{\boldsymbol{L}}\cdot\boldsymbol{\Omega}(\hat{\br}, t)|\psi^{n}_{\bk+\bq'}\rangle = \frac{(2\pi)^3}{N}\sum_{\bK}i\boldsymbol{\nabla}_{\bq'}\delta(\bq'-\bq+\bK)\times
   \intop_{\brho}e^{i(\bq'-\bq)\cdot\brho}\left[\bk u^{m*}_{\bk}(\brho)+i{\boldsymbol\nabla}_{\brho}u^{m*}_{\bk}(\brho)\right]u^{n}_{\bk+\bq'}(\brho)\cdot\boldsymbol{\Omega}(\bq,t) 
\end{equation}  

Since $u^{n}_{\bk+\bq'+\bK}(\brho)=e^{-i\bK\cdot\brho}u^{n}_{\bk+\bq'}(\brho)$, each term in the sum over $\bK$ gives the same contribution and cancels the factor of $N$. Thus, we can safely assume $\bq$ and $\bq'$ to be within the first Brillouin zone and write
\begin{eqnarray}
    \langle\psi^{m}_{\bk}|-\hat{\boldsymbol{L}}\cdot\boldsymbol{\Omega}(\hat{\br}, t)|\psi^{n}_{\bk+\bq'}\rangle= (2\pi)^3i\boldsymbol{\nabla}_{\bq'}\delta(\bq'-\bq)\times  \left<u^{m}_{\bk}|(\bk-i\boldsymbol{\nabla}_{\brho})|u^{n}_{\bk+\bq'}\right>\cdot\boldsymbol{\Omega}(\bq,t)\nonumber\\&&
\end{eqnarray}
Thus, the angular momentum vertex is the prefactor of ${\Omega}(\bq,t)$ as used in Eq. 4 in the main paper.
\begin{equation}
\boldsymbol{L}(\bq';\bq)=i(2\pi)^{3}\boldsymbol{\nabla}_{\bq'}\delta(\bq'-\boldsymbol{q})\times\left( \bk-i\boldsymbol{\nabla_{\rho}}\right)
\end{equation}

We have adopted a notation for $\boldsymbol L(\bq';\bq)$ where the momentum to the right of the semicolon is that of the boson line.

\subsection{Reduction to the continuum} \label{App:continuum}

In an appropriate continuum limit, the matrix elements of the perturbation on a lattice reduce to the appropriate continuum values, i.e., $\bk-i\boldsymbol{\nabla}_{\brho}$ should reduce to the ordinary continuum momentum. Below, we describe how this occurs in the nearly-free-electron limit. Interestingly, such a reduction also occurs in a deep tight-binding limit, which may be more relevant to $d-$ and $f-$electron compounds.

\subsubsection{Nearly free electron limit}

Let us Fourier expand the periodic part of the Bloch wavefunction in terms of the reciprocal lattice vectors, $u_{\bk}^{n}(\brho)=\sum_{\bK}e^{i\bK\cdot\brho}u_{\bk}^{n}(\bK)$. The desired matrix element can then be written as
\begin{equation}
    \left<u^{n}_{\bk}|(\bk-i\boldsymbol{\nabla}_{\brho})|u^{m}_{\bk'}\right>=\sum_{\bK}(\bk+\bK)u^{n*}_{\bk}(\bK)u^{m}_{\bk'}(\bK)
\end{equation}
Now, consider the Hamiltonian for electrons in a periodic potential, $H=-\hbar^2\nabla^2/2m +V(\br)$. For Inserting Bloch wavefunctions $\psi_{\bk}^n(\br)$ into the equation gives
\begin{equation}
    \left[\frac{(\bk-i\boldsymbol{\nabla_{\rho}})^2}{2m} + V(\brho)\right]u_{\bk}^{n}(\brho)=E_{\bk}^{n}u_{\bk}^{n}(\brho)\label{eq: se}
\end{equation}
Fourier expanding $V(\brho)=\sum_{\bK\neq0}V(\bK)e^{i\bK\cdot\brho}$ and integrating over $\brho$ gives a set of linear equations, indexed by $\bK$, for each band:
\begin{eqnarray}
    \frac{(\bk+\bK)^2}{2m}u_{\bk}^{n}(\bK) + \sum_{\bK'}V(\bK-\bK')u_{\bk}^{n}(\bK')&=&E_{\bk}^{n}u_{\bk}^{n}(\bK)\nonumber\\
    \implies u_{\bk}^{n}(\bK) = \frac{\sum_{\bK'}V(\bK-\bK')u_{\bk}^{n}(\bK')}{E^{n}_{\bk}-\frac{(\bk+\bK)^2}{2m}}
    \label{eq:Bloch-Schrodinger}
\end{eqnarray}

Suppose the periodic potential and eigenenergy are weak compared to the free electron kinetic energy at the edge of the first Brillouin zone:
\begin{equation}
    |V(\brho)|,|E_{\bk}^{n}|\ll \frac{\hbar^2(K_\text{min}/2)^2}{2m}\sim\frac{h^2}{8ma^2}
\end{equation}
where $\bK_\text{min}$ is the shortest non-zero reciprocal lattice vector and has length $O(2\pi/a)$ with $a$ being the length scale of the lattice constant. For $\bk\ll \bK_\text{min}$, the denominator of Eq. \ref{eq:Bloch-Schrodinger} is large for any $\bK\neq0$, so the corresponding $u^n_{\bk}(\bK)$ must be small. Specifically,
\begin{equation}
|u_n(\bK\neq0)| \leq \frac{\sum_{\bK'}|V(\bK-\bK')||u_{\bk}^{n}(0)|}{\left| E^{n}_{\bk}-\frac{(\bk+\bK)^2}{2m} \right|} \sim \frac{|V|}{h^2/8ma^2}
\end{equation}
Thus, we can approximate
\begin{equation}
    \left<u^{n}_{\bk}|(\bk-i\boldsymbol{\nabla}_{\brho})|u^{m}_{\bk'}\right> \approx \bk u^{n*}_{\bk}(0)u^{m}_{\bk'}(0)
\end{equation}
thereby recovering the continuum behavior of the momentum operator, $\hat{\boldsymbol{p}}\to\bk$.
\subsubsection{Deep tight-binding limit}

Interestingly, the continuum behavior also arises in a useful opposite ``deep tight-binding" limit. Suppose $V(\brho)$ consists of a sum of Dirac delta function wells within the unit cell:
\begin{equation}
    V(\brho)=\sum_i V_i\delta(\brho-\brho_i)
\end{equation}
For $E<0$ and $1/\sqrt{-2mE}\ll |\brho_i-\brho_j|$ $\forall$ $i,j$, the local spectrum consists of nearly decoupled exponentially decaying waves around each well
\begin{equation}
    \phi_i(\brho) \sim e^{-\sqrt{-2mE}|\brho-\brho_i|}
\end{equation}
and the Bloch functions $u_{\bk}^n(\brho)$ are superpositions of $\phi_i(\brho)$:
\begin{equation}
    u_{\bk}^n(\brho) = \sum_i u_{\bk}^{n,i}\phi_i(\brho)
    \label{eq:u-phi}
\end{equation}
Importantly, for $\bk$ within the first Brillouin zone, the only $\brho$ dependence of $u_{\bk}^n(\brho)$ is through $\phi_i(\brho)$; there are no factors of $e^{i\bK\cdot\brho}$ in the coefficients above. The inner product $\left<u^{n}_{\bk}|i\boldsymbol{\nabla}_{\brho}|u^{m}_{\bk'}\right>$ is now exponentially small as $\left<\phi_i|i\boldsymbol{\nabla}_{\brho}|\phi_j\right>\sim e^{\sqrt{-2mE}|\brho_i-\brho_j|}\ll1$ if $i\neq j$ and $\left<\phi_i|i\boldsymbol{\nabla}_{\brho}|\phi_i\right>=0$ since $\phi_i(\brho)$ have definite parity. Thus, $\hat{\boldsymbol{p}}\to\bk$ is recovered in this limit as well.
Now, we will calculate the response to this rotation using the Kubo formula to first order in $\boldsymbol{\Omega}$. First, we will show the expression for generic band structure in terms of the Berry curvature and then use it to derive explicit results for Weyl fermions.

\section{General results}\label{App:general results}

\subsection{Response function}
Since $\boldsymbol{L}$ violates translational invariance, a spatially monotonic angular velocity $\boldsymbol{\Omega}(\br,t)=e^{-i\bq\cdot\br}\boldsymbol{\Omega}(\bq)$ gives rise to currents $\boldsymbol{j}(\bq')$ with a range of momenta $\bq'$. On the other hand, momentum non-conservation due to $\boldsymbol{L}$ is delicate -- through a $\boldsymbol{\nabla}_{\bq'}\delta(\bq'-\bq)$ function -- so the resultant $\boldsymbol{j}(\bq')$ will still be sharply peaked around $\bq$. Thus, the physical current corresponds to the integrated weight of the current distribution. Physically, the integration captures the fact that momentum-resolved measurement probes necessarily have a non-zero width and, thus, can only detect the integrated weight.

At the level of linear response, the physical current is given by the Kubo formula
\begin{equation}
\left\langle j_{\alpha}(\bq,i\omega_{n})\right\rangle =\intop_{\bq'}\tilde{\chi}_{\alpha\beta}(\bq',\bq,i\omega_{n})\Omega_{\beta}(\bq,i\omega_{n})
\end{equation}
where, $\tilde{\chi}_{\alpha \beta}$ is the response function. 

Depicted by the Feynman diagram Fig. 1(b), it is given at temperature $T$ by
 \begin{eqnarray}
\tilde{\chi}_{\alpha\beta}(\bq',\bq,i\omega_{n})=-T\sum_{i\nu_{n}}\intop_{\bk}\text{tr}[j_{\alpha}(\bk+\bq')G(\bk,i\nu_{n})\nonumber  L_{\beta}(\bq';\bq) \times G(\bk+\bq',i\nu_{n}+i\omega_{n})]    
\end{eqnarray}
where $G(\bk,z)=\left[z-H_{0}(\bk) + i\text{sgn}(\text{Im}z)/2\tau\right]^{-1}$ with a phenomenological quasiparticle relaxation time $\tau$. In the following, we will refer to regimes of relatively small and large $\tau$ as ``dirty" and ``clean" for convenience, which is appropriate for impurity scattering, but stress that $\tau$ is a phenomenological timescale whose origin could be other scattering processes too. As usual, $\boldsymbol{j}(\bk)=\boldsymbol{\nabla}_{\bk} H_0(\bk)$ is the number current; the charge current is $e\boldsymbol{j}(\bk)$.

To proceed, we insert complete sets of states $u^{m}_{\bk}$ and $u_{\bk+\bq'}^{n}$ as written below:
 \begin{eqnarray}
\tilde{\chi}_{\alpha\beta}(\bq',\bq,i\omega_{n})&=&-T\sum_{i\nu_{n}}\intop_{\bk}\langle u^{n}_{\bk+\bq'}|j_{\alpha}(\bk+\bq')G(\bk,i\nu_{n})\sum_{m}|u^{m}_{\bk}\rangle\langle u^{m}_{\bk}|\  L_{\beta}(\bq';\bq) G(\bk+\bq',i\nu_{n}+i\omega_{n})|u^{n}_{\bk+\bq'}\rangle \nonumber \\
&=&-T\sum_{i\nu_{n},n,m}\intop_{\bk}\frac{\langle u^{n}_{\bk+\bq'}|j_{\alpha}(\bk+\bq')|u^{m}_{\bk}\rangle\langle u^{m}_{\bk}|\  L_{\beta}(\bq';\bq) |u^{n}_{\bk+\bq'}\rangle}{\left[i\nu_n-E^m_{\bk}+i\frac{\text{sgn}(\nu_n)}{2\tau}\right]\left[i\nu_n+i\omega_n-E^n_{\bk+\bq}+i\frac{\text{sgn}(\nu_n+\omega_n)}{2\tau}\right]}\nonumber \\
&=&\sum_{n,m}\intop_{\bk}\langle u^{n}_{\bk+\bq'}|j_{\alpha}(\bk+\bq')|u^{m}_{\bk}\rangle\langle u^{m}_{\bk}|\  L_{\beta}(\bq';\bq) |u^{n}_{\bk+\bq'}\rangle S_{nm}(\bk,\bq',i\omega_n)
\end{eqnarray}
where
\begin{equation}
S_{nm}(\bk,\bq',i\omega_n)=-T\sum_{i\nu_n}\frac{1}{\left[i\nu_n-E^m_{\bk}+i\frac{\text{sgn}(\nu_n)}{2\tau}\right]\left[i\nu_n+i\omega_n-E^n_{\bk+\bq'}+i\frac{\text{sgn}(\nu_n+\omega_n)}{2\tau}\right]}
\label{eq:Matsubara}
\end{equation}
is evaluated in the next section. Integrating by parts over $\bq'$ and analytically continuing 
$i\omega_n\to\omega+i0^+$ gives the physical, retarded response function $\chi_{\alpha\beta}(\bq,\omega)=\intop_{\bq'}\tilde{\chi}_{\alpha\beta}(\bq',\bq,i\omega_n\to\omega+i0^+)$:
\begin{eqnarray}
\chi_{\alpha\beta}(\bq,\omega)=\varepsilon^{\beta\mu\nu}\intop_{\bk}\sum_{n,m}i\partial_{q_\nu}[\langle u^{n}_{\bk+\bq}|j_{\alpha}(\bk+\bq)|u^{m}_{\bk}\rangle\ {\langle u^{m}_{\bk}| k_\mu-i\partial_{\rho_\mu}|u^{n}_{\bk+\bq}\rangle}S_{nm}(\bk,\bq,\omega)]
\label{eq:chi-general}
\end{eqnarray}
This is a general expression valid for lattice models at arbitrary $T$, $\bq$, and $\omega$. Its real and imaginary parts define the reactive and dissipative responses, respectively. It can be calculated precisely if the Bloch Hamiltonian $H_0(\bk)$ and the basis states for the Bloch functions [$\phi_i(\brho)$ in Eq. (\ref{eq:u-phi})] are known. The factor $S_{nm}(\bk,\bq,\omega)$ will be calculated generally in the next section.\\

\subsection{Matsubara sum}
We now evaluate the Matsubara sum $S_{nm}(\bk,\bq,i\omega_n)$ in Eq. \ref{eq:Matsubara}. Similar sums appear in textbook calculations of transport. Here, we recap the calculation for completeness and tailor it for the goals of this work.

The summand in Eq. (\ref{eq:Matsubara}) contains no poles but has two branch cuts in the complex frequency plane along $\text{Im}z=0$ and $\text{Im}z=-\omega_n$, where $z$ is the complex generalization of $i\nu_n$. As a result, the Matsubara sum transforms into integrals over the real axis as
\begin{eqnarray}
    S_{nm}(\bk,\bq,i\omega_n)&=&-i\intop_{\nu}f(\nu)\left[\frac{1}{\nu-E_{\bk}^m+\frac{i}{2\tau}}-\frac{1}{\nu-E_{\bk}^m-\frac{i}{2\tau}}\right]\frac{1}{\nu+i\omega_n-E_{\bk+\bq}^n+\frac{i\text{sgn}\omega_n}{2\tau}}\nonumber\\
    &-&i\intop_{\nu}f(\nu)\left[\frac{1}{\nu-E_{\bk+\bq}^n+\frac{i}{2\tau}}-\frac{1}{\nu-E_{\bk+\bq}^n-\frac{i}{2\tau}}\right]\frac{1}{\nu-i\omega_n-E_{\bk}^m-\frac{i\text{sgn}\omega_n}{2\tau}}
\end{eqnarray}
where $f(\nu)=1/(e^{\nu/k_BT}+1)$ is the Fermi function. Upon analytic continuation, $i\omega_n\to\omega+i0^+$,
\begin{equation}
    S_{nm}(\bk,\bq,\omega)=\intop_{\nu}2f(\nu)\left\{\text{Im}\left[\frac{1}{\nu-E_{\bk}^m+\frac{i}{2\tau}}\right]\frac{1}{\nu+\omega-E_{\bk+\bq}^n+\frac{i}{2\tau}}
    +\text{Im}\left[\frac{1}{\nu-E_{\bk+\bq}^n+\frac{i}{2\tau}}\right]\frac{1}{\nu-\omega-E_{\bk}^m-\frac{i}{2\tau}}\right\}
\end{equation}
Shifting $\nu$ by $E^{m}_{\bk}$ and $E^{n}_{\bk+\bq}$ in the two terms and changing $\nu\to-\nu$ in the second term yields
\begin{equation}
    S_{nm}(\bk,\bq,\omega)=\intop_{\nu}2\text{Im}\left[\frac{1}{\nu+\frac{i}{2\tau}}\right]\frac{f(\nu+E^m_{\bk})-f(-\nu+E_{\bk+\bq}^n)}{\nu+E^m_{\bk}-E_{\bk+\bq}^n+\omega+\frac{i}{2\tau}}
    \label{eq:S-general}
\end{equation}
The above expression is valid for general $\bq$, $\omega$, $\tau$, and $T$. At $T=0$, it can be evaluated analytically and gives
 \begin{eqnarray}
    S_{nm}(\bk,\bq,\omega)= \frac{i}{2\pi}\frac{\ln\left[\frac{E_{\bk}^m+\frac{i}{2\tau}}{E^m_{\bk}+\omega+\frac{i}{2\tau}}\frac{E_{\bk+\bq}^n-\frac{i}{2\tau}}{E_{\bk+\bq}^n-\omega-\frac{i}{2\tau}}\right]}{E^m_{\bk}-E_{\bk+\bq}^n+\omega+\frac{i}{\tau}}-\frac{i}{2\pi}\frac{\ln\left[\frac{E_{\bk}^m-\frac{i}{2\tau}}{E^m_{\bk}+\omega+\frac{i}{2\tau}}\frac{E_{\bk+\bq}^n+\frac{i}{2\tau}}{E_{\bk+\bq}^n-\omega-\frac{i}{2\tau}}\right]}{E^m_{\bk}-E_{\bk+\bq}^n+\omega}
    \label{eq:S-T0}
\end{eqnarray}   

\label{general results}
\section{Simplifications and limits}

Ultimately, we will be interested in the limits of $q\to0$ and $\omega\to0$. For small $q$, $S_{nm}(\bk,\bq,\omega)$ for $E^n_{\bk}\neq E^n_{\bk}$ is well-behaved, whereas $S_{nn}(\bk,\bq,\omega)$ and $S_{nm}(\bk,\bq,\omega)$ for $E^n_{\bk}=E^m_{\bk}$ are not. Thus, we can apply $i\partial_{q_\nu}$ on the matrix elements and safely take $q\to0$ in these factors. This yields,
 \begin{eqnarray}
&&\chi_{\alpha\beta}(\bq,\omega)=\nonumber\\&&i\varepsilon^{\beta\mu\nu}\intop_{\bk}\sum_{n,m}\langle \partial_{\nu}u^{n}_{\bk}|j_{\alpha}(\bk)|u^{m}_{\bk}\rangle\  \left(k_\mu\delta^{mn}-\mathcal{A}^{mn}_{\bk\mu}\right)S_{nm}(\bk,\bq,\omega)
+i\varepsilon^{\beta\mu\nu}\intop_{\bk}\sum_{n,m}\langle u^{n}_{\bk}|\partial_{\nu}j_{\alpha}(\bk)|u^{m}_{\bk}\rangle\  \left(k_\mu\delta^{mn}-\mathcal{A}^{mn}_{\bk\mu}\right)S_{nm}(\bk,\bq,\omega)\nonumber\\
&+&i\varepsilon^{\beta\mu\nu}\intop_{\bk}\sum_{n,m}\langle u^{n}_{\bk}|j_{\alpha}(\bk)|u^{m}_{\bk}\rangle\  {\langle u^{m}_{\bk}| k_\mu|\partial_{\nu}u^{n}_{\bk}\rangle}S_{nm}(\bk,\bq,\omega)+i\varepsilon^{\beta\mu\nu}\intop_{\bk}\sum_{n,m}\langle u^{n}_{\bk}|j_{\alpha}(\bk)|u^{m}_{\bk}\rangle\  \left(k_\mu\delta^{mn}-\mathcal{A}^{mn}_{\bk\mu}\right)v_{\bk+\bq \nu}^n\frac{dS_{nm}(\bk,\bq,\omega)}{dE_{\bk+\bq}^n}\nonumber\\&&
\end{eqnarray}   
where $\partial_{\nu}\equiv\partial_{k_\nu}$ and $\mathcal{A}_{\bk\mu}^{mn}=\langle u_{\bk}^m|i\partial_{\rho_\mu}|u_{\bk}^n\rangle$. In the third line, we have used the smoothness of $u_{\bk}^n(\brho)$ in $\bk$ and $\brho$ to drop $\varepsilon^{\beta\mu\nu}\langle u^{m}_{\bk}|i\partial_{\rho_{\mu}}|\partial_{\nu}u^{n}_{\bk}\rangle$. In the fourth line, we have exploited the fact that $S_{nm}(\bk,\bq,\omega)$ depends on $\bq$ only through $E_{\bk+\bq}^n$ We no longer need the basis functions $\phi_i(\brho)$; instead, momentum matrix elements $\mathcal{A}^{mn}_{\bk \mu}$, which are routinely computed by first principles or determined experimentally by measuring optical transitions, suffice. 

Next, we assume a limit where the ``gauge field" $\langle u^{m}_{\bk}|i\partial_{\rho_{\mu}}|u^{n}_{\bk}\rangle$, proportional to the optical matrix element, is negligible, such as the nearly-free-electron or deep tight-binding limits described in Sec. \ref{App:continuum}. Then,
\begin{eqnarray}
\chi_{\alpha\beta}(\bq,\omega)&=&i\varepsilon^{\beta\mu\nu}\intop_{\bk}\sum_{n}\langle \partial_{\nu}u^{n}_{\bk}|j_{\alpha}(\bk)|u^{n}_{\bk}\rangle\  k_\mu S_{nn}(\bk,\bq,\omega)
+ i\varepsilon^{\beta\mu\nu}\intop_{\bk}\sum_{n}\langle u^{n}_{\bk}|\partial_{\nu}j_{\alpha}(\bk)|u^{n}_{\bk}\rangle\  k_\mu S_{nn}(\bk,\bq,\omega)\nonumber\\
&&+ i\varepsilon^{\beta\mu\nu}\intop_{\bk}\sum_{n,m}\langle u^{n}_{\bk}|j_{\alpha}(\bk)|u^{m}_{\bk}\rangle\  {\langle u^{m}_{\bk}|\partial_{\nu}u^{n}_{\bk}\rangle}k_\mu S_{nm}(\bk,\bq,\omega)+ i\varepsilon^{\beta\mu\nu}\intop_{\bk}\sum_{n}v_{\bk \alpha}^n  k_\mu v_{\bk \nu}^n\frac{dS_{nn}(\bk,\bq,\omega)}{dE_{\bk+\bq}^n}
\end{eqnarray}
This further simplifies if we focus on the isotropic part of the response, $\chi_\text{iso}=\frac{1}{3}\sum_\alpha\chi_{\alpha\alpha}$. Verifying that $dS_{nn}(\bk,\bq,\omega)/dE_{\bk+\bq}^n$ is finite, the fourth line does not contribute to $\chi_\text{iso}$ thanks to the factor $\varepsilon^{\alpha\mu\nu}v_{\bk\alpha}^n v_{\bk\nu}^n$. Using
\begin{equation}
    \varepsilon^{\alpha\mu\nu}\partial_\nu\langle u_{\bk}^n|j_\alpha(\bk)|u_{\bk}^n\rangle = \varepsilon^{\alpha\mu\nu}\partial_\nu\partial_\alpha E_{\bk}^n=0
\end{equation}
we get
\begin{eqnarray}
\chi_{\text{iso}}(\bq,\omega)&=&\frac{i\varepsilon^{\alpha\mu\nu}}{3}\intop_{\bk}k_\mu\sum_{n}\langle u^{n}_{\bk}|\left(v_{\bk \alpha}^n-j_{\bk \alpha}\right)|\partial_{\nu}u^{n}_{\bk}\rangle S_{nn}(\bk,\bq,\omega)+\frac{i\varepsilon^{\alpha\mu\nu}}{3}\intop_{\bk}k_\mu\sum_{n\neq m}\langle u^{n}_{\bk}|j_{\alpha}(\bk)|u^{m}_{\bk}\rangle  {\langle u^{m}_{\bk}|\partial_{\nu}u^{n}_{\bk}\rangle}S_{nm}(\bk,\bq,\omega)\nonumber\\
&=&\chi_\text{iso}^\text{intra}+\chi_\text{iso}^\text{inter}
\end{eqnarray}
where we have separated the intraband ($n=m$) and interband ($n\neq m$) contributions. We now study the uniform ($q\to0$ before $\omega\to0$) and static ($\omega\to0$ before $q\to0$) limits for the two types of contributions.
\subsection{Interband, away from band intersections ($E_{\bk}^n\neq E_{\bk}^m$)}

In this case, the uniform and static limits commute and we can set $q=0$ and $\omega=0$ directly. Eq. (\ref{eq:S-T0}) reduces to
\begin{eqnarray}
    S_{nm}(\bk,0,0)&=&-\frac{\arg\left(E_{\bk}^m+\frac{i}{2\tau}\right)-\arg\left(E_{\bk}^n+\frac{i}{2\tau}\right)}{\pi\left(E^m_{\bk}-E_{\bk}^n\right)}\nonumber\\
    &\approx&-\frac{\Theta(-E_{\bk}^m)-\Theta(-E_{\bk}^n)}{E^m_{\bk}-E_{\bk}^n}+\frac{1}{2\pi\tau E^m_{\bk}E_{\bk}^n}\nonumber\\&&
\end{eqnarray}
for $|E_{\bk}^m\tau|,|E_{\bk}^n\tau|\gg1$, where the $\Theta(\dots)$ terms come from branch cuts in $\arg(\dots)$. If one of the energies is zero, then
\begin{equation}
    S_{nm}(\bk,0,0)\approx\begin{cases}
    \frac{1/2-\Theta(-E_{\bk}^m)}{E^m_{\bk}}+\frac{1}{2\pi\tau E_{\bk}^{m2}}&E_{\bk}^n=0,E_{\bk}^m\neq0\\
    \frac{1/2-\Theta(-E_{\bk}^n)}{E^n_{\bk}}-\frac{1}{2\pi\tau E_{\bk}^{n2}}&E_{\bk}^m=0,E_{\bk}^n\neq0
    \end{cases}
\end{equation}
With $\Theta(0)=1/2$, we can compactly write
\begin{equation}
    S_{nm}(\bk,0,0)\approx-\frac{\Theta(-E_{\bk}^m)-\Theta(-E_{\bk}^n)}{E^m_{\bk}-E_{\bk}^n}
\end{equation}
to leading order in $1/\tau E_{\bk}^{m,n}$. Inserting this into the response function gives
\begin{eqnarray}
    \chi_\text{iso}^\text{inter}=-\frac{i\varepsilon^{\alpha\mu\nu}}{3}\intop_{\bk}k_\mu\sum_{n\neq m}\langle u^{n}_{\bk}|j_{\alpha}(\bk)|u^{m}_{\bk}\rangle  {\langle u^{m}_{\bk}|\partial_{\nu}u^{n}_{\bk}\rangle}\frac{\Theta(-E_{\bk}^m)-\Theta(-E_{\bk}^n)}{E^m_{\bk}-E_{\bk}^n}
\end{eqnarray}
as long as there are no band crossings. We can simplify the above by observing that
\begin{eqnarray}
    \langle u_{\bk}^n|\partial_\alpha\left(H_0(\bk)|u_{\bk}^m\rangle\right) &=& \langle u_{\bk}^n|\partial_\alpha\left(E_{\bk}^m|u_{\bk}^m\rangle\right)\nonumber\\
    \implies \langle u^{n}_{\bk}|j_{\alpha}(\bk)|u^{m}_{\bk}\rangle &=& \left(E_{\bk}^m - E_{\bk}^n\right)\langle u_{\bk}^n|\partial_\alpha u_{\bk}^m\rangle + v_{\bk \alpha}^n\delta_{mn}\nonumber\\&&
\end{eqnarray}
and
\begin{equation}
    \langle u_{\bk}^n|\partial_\alpha u_{\bk}^m\rangle + \langle \partial_\alpha u_{\bk}^n|u_{\bk}^m\rangle = \partial_\alpha \left<u_{\bk}^n|u_{\bk}^m\right> = 0
\end{equation}
Thus,
\begin{eqnarray}
    \chi_\text{iso}^\text{inter}&=&-\frac{i\varepsilon^{\alpha\mu\nu}}{3}\intop_{\bk}k_\mu\sum_{n\neq m}\langle u_{\bk}^n|\partial_\alpha u_{\bk}^m\rangle{\langle u^{m}_{\bk}|\partial_{\nu}u^{n}_{\bk}\rangle}\left[\Theta(-E_{\bk}^m)-\Theta(-E_{\bk}^n)\right]\nonumber\\
    &=&\frac{i\varepsilon^{\alpha\mu\nu}}{3}\intop_{\bk}k_\mu\sum_{n\neq m}\left\{\langle \partial_\alpha u_{\bk}^n| u_{\bk}^m\rangle{\langle u^{m}_{\bk}|\partial_{\nu}u^{n}_{\bk}\rangle}-{\langle \partial_\nu u^{n}_{\bk}|u^{m}_{\bk}\rangle}\langle u_{\bk}^m|\partial_\alpha u_{\bk}^n\rangle\right\}\Theta(-E_{\bk}^n)\nonumber\\
    &=&\frac{i\varepsilon^{\alpha\mu\nu}}{3}\intop_{\bk}k_\mu\sum_{n}\left\{\langle \partial_\alpha u_{\bk}^n| \partial_{\nu}u^{n}_{\bk}\rangle-\langle \partial_\nu u^{n}_{\bk}|\partial_\alpha u_{\bk}^n\rangle\right\}\Theta(-E_{\bk}^n)\nonumber\\
    &=&-\frac{2}{3}\intop_{\bk}\sum_n \Theta(-E_{\bk}^n)\bk\cdot\boldsymbol{F}_{\bk}^n
    \label{eq:chi-inter}
\end{eqnarray}
where $\boldsymbol{F}_{\bk}^{n}=i\left\langle \boldsymbol{\nabla}_{\bk}u_{\bk}^{n}\left|\times\right|\boldsymbol{\nabla}_{\bk}u_{\bk}^{n}\right\rangle $
is the Berry curvature of the $n^{th}$ band. Going from the second to the third line, we added the $m=n$ term to the summation, noting that it vanishes anyway. Thus, $\chi_\text{iso}^\text{inter}$ in continuum Hamiltonians is given by the first moment of the Berry curvature of the occupied Bloch states if there are no band intersections. We will see shortly that the above result holds even when band intersections are included as long as the intersections are not within $O(1/\tau)$ of the Fermi level.
\subsection{Intraband or at a band intersection ($E_{\bk}^n= E_{\bk}^m$)}

In this case, the uniform and static limits do not commute, so we must consider them separately. We assume that band intersections, if any, are not close to the Fermi level. If they are, Bloch functions become non-differentiable at these points, and the formalism we have developed so far breaks down. Physically, tuning the Fermi level to a band intersection invalidates smooth, semiclassical nature of the $q\to0$ limit and demands a non-perturbative, fully quantum calculation.
\subsubsection{Uniform limit: $q\to0$ before $\omega\to0$}
At $q=0$ and $E_{\bk}^n= E_{\bk}^m$, Eq. (\ref{eq:S-T0}) becomes
\begin{eqnarray}
    S_{nm}(\bk,0,\omega)=\frac{i}{2\pi}\ln\left[\frac{E_{\bk}^m+\frac{i}{2\tau}}{E_{\bk}^m-\omega-\frac{i}{2\tau}}\frac{E_{\bk}^m-\frac{i}{2\tau}}{E^m_{\bk}+\omega+\frac{i}{2\tau}}\right]\frac{1}{\omega(i\omega\tau-1)}
\end{eqnarray}
For $E_{\bk}^m\neq0$, which is true away from the Fermi level, we consider the low-frequency regime, $|\omega|\ll\left|E_{\bk}^m\right|$ and approximate
\begin{equation}
S_{nm}(\bk,0,\omega)=\frac{1}{\pi}\text{Im}\left[\frac{1}{E_{\bk}^m+\frac{i}{2\tau}}\right]\frac{1}{i\omega\tau-1}
\end{equation}
Now taking $\omega\to0$ in the two extremes $\omega\tau\to\infty$ (``clean") and $\omega\tau\to0$ (``dirty"),
\begin{equation}
    S_{nm}(\bk,0,\omega\to0)=\begin{cases}
        -\frac{\delta(E_{\bk}^m)}{i\omega\tau}\to0&|\omega\tau|\gg1\\
        -\frac{1}{\pi}\text{Im}\left[\frac{1}{E_{\bk}^m+\frac{i}{2\tau}}\right]&|\omega\tau|\ll1\\
    \end{cases}
    \label{eq:S-intra-uni-nonzeroE}
\end{equation}
On the other hand, if $E_{\bk}^m=0$,
\begin{equation}
    S_{nm}(\bk,0,\omega)=-\frac{i\ln(1-2i\tau\omega)}{\pi\omega(i\omega\tau-1)}
\end{equation}
As $\omega\to0$,
\begin{equation}
    S_{nm}(\bk,0,\omega)=\begin{cases}
        0&|\omega\tau|\gg1\\
        \frac{2\tau}{\pi}&|\omega\tau|\ll1
    \end{cases}
\end{equation}
which is equivalent to Eq. (\ref{eq:S-intra-uni-nonzeroE}) at $E_{\bk}^m=0$. In other words, the $\omega\to0$ and the $E_{\bk}^m\to0$ limits commute. Thus, we can compactly write
\begin{equation}
    S_{nm}(\bk,0,\omega\to0)=\begin{cases}
        0&|\omega\tau|\gg1\\
        -\frac{1}{\pi}\text{Im}\left[\frac{1}{E_{\bk}^m+\frac{i}{2\tau}}\right]&|\omega\tau|\ll1
    \end{cases}
    \label{eq:S-uni}
\end{equation}
for all $E_{\bk}^m$. Thus, the $|\omega\tau|\gg1$ regime gives no additional contribution to $\chi_\text{iso}^\text{inter}$ even if there are band crossings.

If $|\omega\tau|\ll1$, $\chi_\text{iso}^\text{inter}$ receives an additional contribution from regions where $E_{\bk}^m=E_{\bk}^n$:
\begin{eqnarray}
    \Delta\chi_\text{iso}^\text{inter}=-\frac{i\varepsilon^{\alpha\mu\nu}}{3\pi}\intop_{\bk}k_\mu\sum_{n\neq m;E_{\bk}^m=E_{\bk}^n}\langle u^{n}_{\bk}|j_{\alpha}(\bk)|u^{m}_{\bk}\rangle  {\langle u^{m}_{\bk}|\partial_{\nu}u^{n}_{\bk}\rangle}\text{Im}\left[\frac{1}{E_{\bk}^m+\frac{i}{2\tau}}\right]
\end{eqnarray}
The factor $\text{Im}\left[\frac{1}{E_{\bk}^m+\frac{i}{2\tau}}\right]$ stipulates that the above correction is governed by band intersections within $O(1/\tau)$ of the Fermi level -- a situation we excluded from the outset in this section -- so the correction will be negligible.

In the uniform limit, $\chi_\text{iso}^\text{intra}$ vanishes if $|\omega\tau|\gg1$ according to Eq. (\ref{eq:S-uni}). If $|\omega\tau|\ll1$,
\begin{eqnarray}
    \chi_\text{iso}^\text{intra} &=&-\frac{i\varepsilon^{\alpha\mu\nu}}{3\pi}\intop_{\bk}k_\mu\sum_{n}\langle u^{n}_{\bk}|\left(v_{\bk \alpha}^n-j_{\bk \alpha}\right)|\partial_{\nu}u^{n}_{\bk}\rangle\text{Im}\left[\frac{1}{E_{\bk}^n+\frac{i}{2\tau}}\right]\nonumber\\
    &=&-\frac{i\varepsilon^{\alpha\mu\nu}}{3\pi}\intop_{\bk}k_\mu\sum_{n} \langle \partial_\alpha u_{\bk}^n|\left(H-E_{\bk}^n\right)|\partial_\nu u_{\bk}^n\rangle\text{Im}\left[\frac{1}{E_{\bk}^n+\frac{i}{2\tau}}\right]\nonumber\\
    &=&\frac{2}{3}\intop_{\bk}\sum_{n} \frac{\boldsymbol{m}_{\bk}^n\cdot\bk}{e}\frac{1}{\pi}\text{Im}\left[\frac{1}{E_{\bk}^n+\frac{i}{2\tau}}\right]
\end{eqnarray}
where $\boldsymbol{m}^n_{\bk} = \frac{ie}{2}\langle\boldsymbol{\nabla}u^n_{\bk}|\times(H_{\bk} - E^n_{\bk})|\boldsymbol{\nabla}u^n_{\bk}\rangle$ is the orbital moment of the Bloch state $|u_{\bk}^n\rangle$. To leading order in $1/\tau$, 
\begin{equation}
    \chi_\text{iso}^\text{intra}=-\frac{2}{3}\intop_{\bk}\sum_{n} \frac{\boldsymbol{m}_{\bk}^n\cdot\bk}{e}\delta(E_{\bk}^n)
\end{equation}
Thus, the contribution is given by the orbital magnetic moment integrated over the Fermi surface.

\subsubsection{Static limit: $\omega\to0$ before $q\to0$}

At $\omega=0$,
\begin{equation}
    S_{nm}(\bk,\bq,0)=-\frac{\arg\left(E_{\bk}^m+\frac{i}{2\tau}\right)-\arg\left(E_{\bk+\bq}^n+\frac{i}{2\tau}\right)}{\pi\left(E^m_{\bk}-E_{\bk+\bq}^n\right)}
\end{equation}
Now taking $q\to0$, we find
\begin{equation}
    S_{nm}(\bk,\bq\to0,0)=\begin{cases}
        \delta(E_{\bk}^m)&|\boldsymbol{v}_{\bk}^n\cdot\bq|\tau\gg1\\
        -\frac{1}{\pi}\text{Im}\left[\frac{1}{E^m_{\bk}+\frac{i}{2\tau}}\right]&|\boldsymbol{v}_{\bk}^n\cdot\bq|\tau\ll1
    \end{cases}
\end{equation}
Inserting this into $\chi_\text{iso}^\text{inter}$ for $|\boldsymbol{v}_{\bk}^n\cdot\bq|\tau\gg1$ (``clean" limit)  gives a correction to $\chi_\text{iso}^\text{inter}$ from band intersections at or near the Fermi level. Since we have excluded such situations, the correction vanishes for $|\boldsymbol{v}_{\bk}^n\cdot\bq|\tau\gg1$ and is negligible for $|\boldsymbol{v}_{\bk}^n\cdot\bq|\tau\ll1$.

On the other hand, the intraband susceptibility for $|\boldsymbol{v}_{\bk}^n\cdot\bq|\tau\gg1$ (``dirty" limit) is given by
\begin{eqnarray}
    \chi_\text{iso}^\text{intra} &=& \frac{i\varepsilon^{\alpha\mu\nu}}{3}\intop_{\bk}k_\mu\sum_{n}\langle u^{n}_{\bk}|\left(v_{\bk \alpha}^n-j_{\bk \alpha}\right)|\partial_{\nu}u^{n}_{\bk}\rangle \delta(E_{\bk}^n)\nonumber\\
    &=&-\frac{2}{3}\intop_{\bk}\sum_{n} \frac{\boldsymbol{m}_{\bk}^n\cdot\bk}{e}\delta(E_{\bk}^n)
\end{eqnarray}
identical to the result obtained in the uniform limit $q=0$, $\omega\to0$, $\tau\to\infty$ with $|\omega\tau|\ll1$. When $|\boldsymbol{v}_{\bk}^n\cdot\bq|\tau\ll1$, the result to leading order in $1/\tau$ is
\begin{equation}
    \chi_\text{iso}^\text{intra}=-\frac{2}{3}\intop_{\bk}\sum_{n} \frac{\boldsymbol{m}_{\bk}^n\cdot\bk}{e}\delta(E_{\bk}^n)
\end{equation}
identical to the expression in the dirty limit.

\twocolumngrid

\bibliographystyle{apsrev4-2}
\bibliography{library}
\end{document}